\documentclass[letterpaper,aps,pra,twocolumn,english,showpacs,superscriptaddress]{revtex4-1}

\usepackage{tikz}
\usepackage[T1]{fontenc}
\usepackage[latin1]{inputenc}
\usepackage{amssymb, amsmath, amsfonts, bm, bbm}
\usetikzlibrary{decorations.pathmorphing}
\usetikzlibrary{decorations.pathreplacing}
\usetikzlibrary{shapes,arrows,snakes}
\usepackage{natbib}
\usepackage[normalem]{ulem} 

\providecommand{\LyX}{L\kern-.1667em\lower.25em\hbox{Y}\kern-.125emX\@}

\definecolor{dgreen}{RGB}{50,128,128}
\usepackage{multirow}
\usepackage{graphicx}
\input{epsf}
\usepackage{babel}
\usepackage[pdftex]{hyperref}
\hypersetup{
    colorlinks,%
    citecolor=black,%
    filecolor=black,%
    linkcolor=black,%
    urlcolor=black     
}

\newcommand{\ket}[1]{|#1\rangle}
\newcommand{\bra}[1]{\langle #1|}
\newcommand{\mbf}[1]{\mathbf{#1}}
\newcommand{\lae}{\lower 2pt \hbox{$\, \buildrel {\scriptstyle <}\over {\scriptstyle\sim}\,$}}

\begin{document}
\title{Robust quantum logic in neutral atoms via adiabatic Rydberg dressing}

\author{Tyler Keating}
\affiliation{Center for Quantum Information and Control (CQuIC), University of New Mexico, Albuquerque NM 87131}
\affiliation{Department of Physics and Astronomy, University of New Mexico,
Albuquerque NM 87131}

\author{Robert L. Cook}
\affiliation{Center for Quantum Information and Control (CQuIC), University of New Mexico, Albuquerque NM 87131}
\affiliation{Department of Physics and Astronomy, University of New Mexico, Albuquerque NM 87131}

\author{Aaron Hankin}
\affiliation{Center for Quantum Information and Control (CQuIC), University of New Mexico, Albuquerque NM 87131}
\affiliation{Sandia National Laboratories, Albuquerque, NM 87185}

\author{Yuan-Yu Jau}
\affiliation{Center for Quantum Information and Control (CQuIC), University of New Mexico, Albuquerque NM 87131}
\affiliation{Sandia National Laboratories, Albuquerque, NM 87185}

\author{Grant W. Biedermann}
\affiliation{Center for Quantum Information and Control (CQuIC), University of New Mexico, Albuquerque NM 87131}
\affiliation{Department of Physics and Astronomy, University of New Mexico, Albuquerque NM 87131}
\affiliation{Sandia National Laboratories, Albuquerque, NM 87185}

\author{Ivan H. Deutsch}
\affiliation{Center for Quantum Information and Control (CQuIC), University of New Mexico, Albuquerque NM 87131}
\affiliation{Department of Physics and Astronomy, University of New Mexico, Albuquerque NM 87131}

\begin{abstract}
We study a scheme for implementing a controlled-Z (CZ) gate between two neutral-atom qubits based on the Rydberg blockade mechanism in a manner that is robust to errors caused by atomic motion.  By employing adiabatic dressing of the ground electronic state, we can protect the gate from decoherence due to random phase errors that typically arise because of atomic thermal motion.  In addition, the adiabatic protocol allows for a Doppler-free configuration that involves counterpropagating lasers in a $\sigma_+/\sigma_-$ orthogonal polarization geometry that further reduces motional errors due to Doppler shifts.  The residual motional error is dominated by dipole-dipole forces acting on doubly-excited Rydberg atoms when the blockade is imperfect. For reasonable parameters, with qubits encoded into the clock states of $^{133}$Cs, we predict that our protocol could produce a CZ gate in $<10$ $\mu$s with error probability on the order of $10^{-3}$.
\end{abstract}
\pacs{34.50.-s,34.10.+x}
\maketitle

\section{Introduction}
A primary obstacle to scalable quantum computation is the requirement that qubits must interact strongly with each other to produce entangling gates and conditional logic, while interacting weakly with their environment to minimize decoherence.  Neutral atom qubits are naturally well-isolated from their environments, but their interactions with each other tend to be similarly weak.  As shown in the seminal work of Jaksch {\em et al.}~\cite{jaksch2000}, one way around this difficulty is to couple ground-state neutral atoms to highly excited Rydberg states, producing strong dipole-dipole interactions on demand while preserving the robustness properties in between operations. In particular, the Rydberg blockade~\cite{lukin2001} provides a direct mechanism for producing entangling interactions between atoms on demand between individually trapped atoms \cite{browaeys, saffmanentangled, hankin}.

There have been numerous proposals to use the Rydberg blockade as a mechanism for implementing two-qubit quantum logic gates  \cite{saffmanreview}, and experimental progress in producing  a controlled-NOT (CNOT)  gate has been  promising \cite{demonstration, saffmanentangled}.  In the standard approach of fast gates, one employs short resonant pulses, in conjunction with the Rydberg blockade to induce the requisite entangling interaction.  However, such a mechanism is not robust to thermal motion of the atoms, which imparts random phases on the two-atom state that vary from shot to shot.  Indeed, such random phases are impediments to the direct observation of entanglement in the signature two-atom Rydberg blockade~\cite{browaeys}.  More generally, the decoherence arising from coupling internal (electronic) and external (motional) degrees of freedom  is a dominant source of error that limits the implementation of high-fidelity quantum gates~\cite{saffman_experimental}.

To address this issue we propose a method of implementing entangling gates that is robust to errors caused by atomic motion by {\em dressing} the ground states via the Rydberg blockade~\cite{Molmer2002, Rolston, Zoller2010, Pohl2014, Pfau2014}, and evolving the system {\em adiabatically}.   The original proposal of Jaksch {\em et al.}~\cite{jaksch2000} examined adiabatic evolution as a mechanism for relaxing the requirement of single atom addressability, and only did so for atoms cooled to the ground state of motion.  Subsequent proposals have suggested various modifications, but most  either ignore thermal motion in order to focus on electronic effects~\cite{molmer_adiabatic, decoherence_free} or require experimental parameters that are challenging to achieve~\cite{molmer_weak_dressing}.  Our motivation is to use adiabaticity to substantially improve the robustness to errors caused by atomic motion, and thereby achieve high-fidelity operation with current technology.  Adiabatic evolution, a well known strategy for suppressing certain error mechanisms, is a paradigm for implementation of a quantum algorithm~\cite{farhi}, and we have previously studied this in the context of the Rydberg blockade~\cite{rydberg_aqc}.  Similar robustness was recently studied in adiabatic passage of atoms to a doubly-excited Rydberg state~\cite{Molmer2014}, which might be used as a mechanism to generate quantum logic gates.

Adiabatic evolution does not protect against all types of decoherence, however, and the motional errors we consider are not strongly suppressed by adiabaticity alone. In fact, motional errors have been among the main fidelity-limiting factors in recent attempts to produce an adiabatic gate~\cite{calarco}.  The protocol we consider is compatible with a ``Doppler-free'' laser configuration, in which the qubits are excited by two counterpropagating beams rather than just a single beam. Such a configuration does not directly reduce the terms in the Hamiltonian that lead to motional decoherence, but it changes their form to one more amenable to adiabatic suppression. Taken together, adiabatic dressing and a Doppler-free configuration produce more than an order-of-magnitude reduction of motional decoherence that neither change achieves on its own.

The remainder of this article is organized as follows. In Sec. II we describe the dressed Rydberg blockade and describe a protocol to perform a controlled-Z gate adiabatically via this interaction. In Sec. III we examine the errors arising from atomic motion, including both single-qubit errors due to thermal motion and two-qubit errors due to imperfect Rydberg blockade. Among these errors, we identify the Doppler shift as a primary obstacle to achieving high gate fidelities, and show how it can be suppressed using a Doppler-free configuration with counterpropagating lasers. In Sec. IV, we numerically simulate the performance of such a gate with realistic experimental parameters and find that error probabilities on the order of $10^{-3}$ should be possible. Finally, we offer some concluding remarks in Sec. V.

\section{Implementing a CZ gate}
\subsection{The Dressed-Blockade Interaction}
For concreteness, we consider qubits encoded in single $^{133}$Cs atoms, individually trapped in tightly focused optical tweezers, with a typical separation of 5-10 microns (see Fig. 1).   Qubits are encoded in the magnetically insensitive ``clock'' states, $\ket{0}\equiv\ket{6S_{1/2}; F=4,M_F=0}$ and $\ket{1} \equiv \ket{6S_{1/2}; F=3,M_F=0}$.  We consider direct excitation to a high-lying Rydberg level, $\ket{r} \equiv \ket{84P_{3/2}; M_J}$ by a single exciting laser at $\lambda_L \approx 319$ nm  in the absence of the trap which is turned off during the duration of the interaction so the atoms undergo ballistic motion~\cite{hankin}.  In the absence of the dipole-dipole interaction, each atom (labeled $i = a,b$) interacts with a laser propagating on the  interatomic $z$ axis.  The  Hamiltonian individually governing the dynamics of the two atoms is (in the two-level, rotating wave approximation, $\hbar=1$),
\begin{equation}\label{eq:1atomH}
H_i = \frac{p_i^2}{2m} - \Delta \ket{r}_i\bra{r} + \frac{\Omega}{2}( e^{i k_L z_i}\ket{r}_i\bra{0}+e^{-i k_L  z_i}\ket{0}_i\bra{r}).
\end{equation}
When including the dipole-dipole interaction of atoms in the Rydberg states, the two-atom Hamiltonian takes the form,
\begin{equation}\label{eq:2atomH}
H = H_a\otimes\mathbbm{1} + \mathbbm{1}\otimes H_b + V_{dd}(z_b - z_a) \ket{rr}\bra{rr},
\end{equation}
where $V_{dd}(z)$ is the dipole-dipole potential for two atoms excited to the Rydberg state. This form of the interaction energy is approximately correct for atoms separated by a large enough distance such that the interaction is perturbative when compared to the splitting of the atomic Rydberg levels (e.g., in the van der Waals regime).  For more closely spaced atoms, the electrostatic forces will strongly mix many atomic orbitals into molecular-type orbitals, so that the double excitation is no longer of the form $\ket{rr}\bra{rr}$, for a single Rydberg level \cite{Shaffer2006}.  Nevertheless, as long as the blockade is strong, we can obtain the essential physics by considering only one doubly-excited state with a given dipole-dipole potential.

\begin{figure*}
[t]\resizebox{12cm}{!}
{\includegraphics{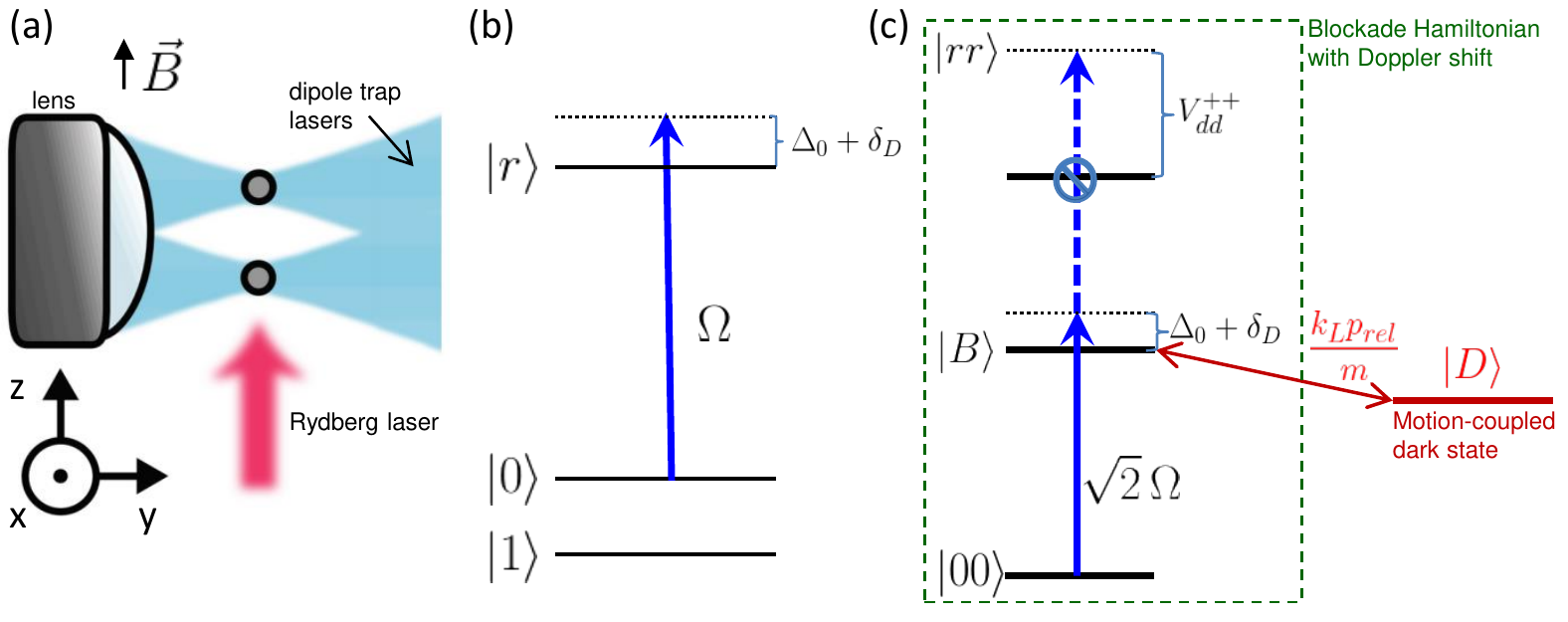}}
\caption{ \label{fig:level_diagram} (Color online) (a) Schematic for the CPHASE gate. Two cesium atoms are trapped and cooled in dipole traps, several $\mu$m apart. During the CPHASE gate, the trapping lasers are turned off and the atoms are illuminated by a 319 nm Rydberg laser. A bias magnetic field ensures that the laser's propagation axis coincides with the atomic quantization axis. (b) In each atom, the logical-$\ket{0}$ state is coupled to a $\ket{84P_{3/2};M_J}$ Rydberg state. The coupling laser has Rabi rate $\Omega$ and detuning from atomic resonance $\Delta_0$, with a momentum-dependent Doppler shift $\delta_D\equiv k_lp/m$. (c) In the two-atom basis, $\ket{00}$ is coupled to the bright state $\ket{B}$, again with base detuning $\Delta_0$ and Doppler shift $\delta_D$. Excitation to $\ket{rr}$ is blockaded by the dipole-dipole interaction $V_{dd}$. Atomic motion further couples $\ket{B}$ to a dark state, $\ket{D}$, outside the ideal blockade subspace.}
\end{figure*}

The position dependent phases $\exp(\pm i k_L z_i)$ associated with photon recoil can be removed from the Hamiltonian by moving to a frame where a Rydberg excited atom is moving with a velocity $\boldsymbol{v} = - \mbf{k}_L/m$ with respect to the lab frame, yielding,
\begin{equation}\label{eq:1atomHmoving}
   H_i \Rightarrow \frac{p_i^2}{2m} - \left(\Delta - \frac{k_L p_i}{m} \right)\ket{r}_i\bra{r} + \frac{\Omega}{2}( \ket{r}_i\bra{0} + \ket{0}_i\bra{r}).
\end{equation}
Here we have absorbed the constant recoil energy into the standard definition of the detuning, $\Delta \rightarrow \Delta - k^2_L/2m$, and see explicitly the Doppler shift, $k_L p_i/m$.  The single atom laser induced light shift (LS) on the ground state at zero momentum is  $\Delta E_{LS}^{(1)}= \frac{1}{2}\left(-\Delta + \operatorname{sign}(\Delta) \sqrt{\Delta^2+\Omega^2}\right)$ 

As the interaction is only a function of the relative atomic distance, it is useful to express the Hamiltonian in terms of the center-of-mass $\mbf{P}_{cm} = \mbf{p}_a +\mbf{p}_b$ and relative $\mbf{p}_{rel} = (\mbf{p}_b - \mbf{p}_a)/2$ momentum coordinates. In addition, the laser field only couples the logical state $\ket{0 0}$ to a symmetric superposition of one excited and one ground state atom.  Defining the bright and dark states of this two-atom system, $ \ket{B} \equiv \left( \ket{r0} + \ket{0r} \right)/\sqrt{2}$ and  $\ket{D} \equiv \left( \ket{r0} - \ket{0r} \right)/\sqrt{2}, $   Eq. (\ref{eq:2atomH}) can be rewritten as  
\begin{equation} \label{perturbedH}
\begin{split}
H & \approx H_0 + H_1, \\
H_0 &=   - \Delta \Big(\ket{B}\bra{B} + \ket{D}\bra{D}  \Big)  -\Big(2 \Delta- V_{dd}(\bar{z}) \Big)  \ket{rr}\bra{rr} \\
&+\frac{\sqrt{2} \Omega}{2} \Big( \ket{B}\bra{00} + \ket{00}\bra{B} +  \ket{rr}\bra{B} + \ket{B}\bra{rr} \Big), \\
H_1\, & = \,  T +V_{grad}+ V_{Dop}.
\end{split}
\end{equation}
Written in this form, $H_0$ is the ``frozen atom'' model including only the internal state dynamics, that show the usual $\sqrt{2} \Omega$ Rabi flopping between the double-ground $\ket{00}$, single-Rydberg bright $\ket{B}$, and double Rydberg  $\ket{rr}$ states.  The blockade energy is taken at the mean atomic separation $\bar{z}$. $H_1$ accounts for the effects of atomic motion according to
\begin{equation} \label{motionalVs}
\begin{split}
T &\equiv \frac{P_{cm}^2}{4m} + \frac{p_{rel}^2}{m},\\
 V_{grad} &\equiv \frac{dV_{dd}}{dz}\Big|_{\bar{z}} (z - \bar{z})\,\ket{rr}\bra{rr},\, \, \text{and,}\\
  V_{Dop} &\equiv \frac{k_L P_{cm}}{2m} \Big(\ket{B}\bra{B} + \ket{D}\bra{D} + 2 \ket{rr}\bra{rr} \Big)\\
 &\quad- \frac{k_L p_{rel}}{m} \Big(\ket{B}\bra{D}+\ket{D}\bra{B}\Big).
\end{split}
\end{equation}
$T$ is the kinetic energy; this term does not entangle internal and external degrees of freedom and thus is unimportant in the perturbation to the logic gate. $V_{grad}$ accounts for the interatomic forces due to the local gradient of the dipole-dipole potential for the doubly-excited Rydberg state and results from linearizing $V_{dd}$ about the point $z = \bar{z}$.   $V_{Dop}$ describes the effect of the Doppler shift.  This includes a term diagonal in the $\{\ket{B},\ket{D}\}$ basis that depends on the center of mass momentum. The off-diagonal terms in $V_{Dop}$ account for the coupling between bright and dark states due to the relative motion of the atoms, familiar in studies of coherent population trapping~\cite{aspect}.  This term leads to random phases induced by thermal motion that cause errors and reduce the entangling action of the interaction.

The eigenstates of $H_0$ are completely decoupled from the motional degrees of freedom and define the adiabatic basis.  The problem can be simply diagonalized; the general case has been studied in~\cite{Rolston}.  In a strongly blockaded regime, $|V_{dd}(\bar{z})| \gg |\Delta|, \Omega$, excitation to the doubly-excited state $\ket{ rr }$ is suppressed by a factor of order $(V_{dd})^2/(\Omega^2 + \Delta^2)$.  The ground state $\ket{00}$ and the entangled bright state $\ket{B}$ form an effective two-level system, and coupling to $\ket{rr}$ can be treated as a perturbation.  The two-atom ground-state light-shift energy is then approximately,  $E^{(2)}_{LS} \approx \frac{1}{2}\left(-\Delta
 + \operatorname{sign}(\Delta) \sqrt{\Delta^2+2\Omega^2}\right)$~\cite{Rolston}.  The effective atomic interaction strength $J$ is the difference between the two-atom light shift and that for two atoms in the absence of the dipole-dipole force, $J\equiv E_{LS}^{(2)}-2E_{LS}^{(1)}$.  For weak dressing, $\Omega \ll |\Delta|$, $J \approx -\Omega^4/(8\Delta^3)$. As we will see, however, the regime of the highest fidelity operation occurs for strong dressing, close to equal superpositions of ground and bright states.  In our previous analysis, we found $J/2\pi = 500$ kHz to be experimentally feasible \cite{rydberg_aqc}.

\subsection{The CZ Gate Protocol}
Given an interaction of this form, it is straightforward to produce a two-qubit logic gate in a manner analogous to Jaksch \emph{et al.}~\cite{jaksch2000}.  Adiabatically increasing the Rydberg laser power while decreasing the detuning creates the coupling, $J(t)$.  Concurrently, the instantaneous ground state of $H_0$ evolves from the bare $\ket{00}$ state into a ``dressed'' state with some admixture of Rydberg character, $\ket{\widetilde{00}}=c_0\ket{00}+c_B\ket{B}+c_{rr}\ket{rr}$, where the coefficients $c_0$, $c_B$, and $c_r$ depend on  the time-dependent parameters $\Delta(t)$ and $\Omega(t)$, as well as the static blockade $V_{dd}(\bar{z})$.   Perfect adiabatic state transfer is ensured by satisfying the adiabatic condition, $|\bra{e}\frac{d}{dt}H_0\ket{\widetilde{00}}| \ll | E(e)-E(\widetilde{00})|^2$, where $\ket{e}$ is any one of the instantaneous excited states of $H_0$. Inverting this ramp returns the system to the bare logical subspace, with the addition of nontrivial phases. When the adiabatic condition is satisfied, $J(t)$ is the rate at which the dressed ground state accumulates the entangling phase.  Integrating the evolution over the total time duration of the gate, $[0, T]$, gives a unitary map, $U^{(2)}_{LS}$, that, when restricted to the two-qubit-logical subspace, takes the diagonal form,
\begin{equation}
\begin{split}
U^{(2)}_{LS}&=\sum_{xy=0,1} e^{-i\phi_{xy}}\ket{xy}\bra{xy}, \,\,  \rm{where},\\
\phi_{11} &=0; \, \, \phi_{10}=\phi_{01} =\int_0^T dt\, E^{(1)}_{LS}(t); \, \, \phi_{00} =\int_0^T dt\, E^{(2)}_{LS}(t). \end{split}
\end{equation}
Following this with the inverse of local single qubit unitaries, $U^{(1)}_{LS} = \exp( -i \phi_{10} \ket{0}\bra{0} )$, cancels the single atom light shifts, yielding the controlled phase gate, $U_{C\phi_J}$,
\begin{equation}
\begin{split}
U_{C\phi_J} &=\big( U^{(1)}_{LS} \otimes U^{(1)}_{LS})^\dag\, U^{(2)}_{LS} = e^{-i \phi_{J} \ket{00}\bra{00}},\\
&{\rm where}\ \phi_{J} =\int_0^T dt\, J(t).
\end{split}
\end{equation}
The single-atom light shifts can be compensated by, e.g., applying microwave pulses or Raman lasers.  The case where $\phi_J=\pi$ is of particular interest, since $U_{C\pi} \equiv U_{CZ}$  is the controlled-Z (CZ) gate, which, up to local unitaries, is equivalent to a controlled-X (CX, or CNOT) gate.

The speed of the gate is set by balancing the requirements that one adiabatically follows the dressed ground state of the Hamiltonian during the implementation of the gate while avoiding the errors that accumulate over time.  One fundamental source of such errors is the finite lifetime of the Rydberg state, $\Gamma^{-1}$.  Decay of $\ket{r}$ will not only dephase  the qubits, but with high probability optically pump them into magnetic sublevels outside the computational space, so we treat this as loss. This effect can be described as the action of a non-Hermitian, effective Hamiltonian with an imaginary part to the detuning: $\Delta \rightarrow \Delta - i \Gamma/2$. Over the full duration $T$ of a gate, such loss will reduce the trace of the density matrix. For a large detuning, the interaction strength scales as $J \sim -\Omega^4/\Delta^3$, while the decay rate due to absorption of a photon and decay of the Rydberg state scales as $\gamma \sim \Omega^2 \Gamma/\Delta^2$. This implies that it is not advantageous to remain in the large detuning limit, but to instead adiabatically sweep to resonance, where the dressing is maximum, while simultaneously avoiding, to the maximum degree possible, double excitation of two atoms into the Rydberg state.

\begin{figure}
\includegraphics{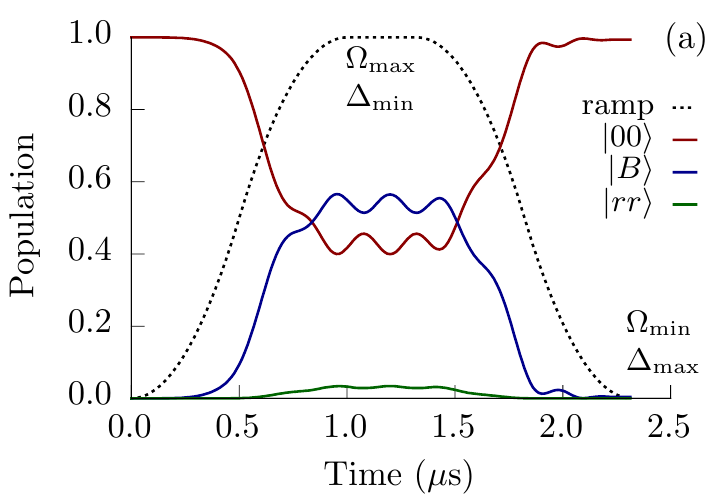}
\caption{\label{fig:pulseshape}(Color online) Pulse shape and bare state populations over the course of a gate with experimentally feasible parameters: pulse rise time 1 $\mu$s, Rabi frequency sweep $\Omega/2\pi =0 \rightarrow$ 3 MHz, detuning sweep $\Delta/2\pi = 6 \rightarrow$ 0 MHz,  Rydberg decay rate $\Gamma/2\pi $= 3.7 kHz (blackbody limited lifetime), and interatomic separation $r=5$ $\mu$m. As the laser turns on and is tuned to resonance, the bare ground state (red) is dressed by admixing significant bright state (blue) population, while the blockaded $\ket{rr}$ state (green) remains mostly unpopulated. Adiabaticity and available interaction strength set comparable constraints in this case, so that the laser pulse shape that best achieves the desired evolution is neither square-topped nor triangular.}
\end{figure}

The shape of the laser pulse can strongly influence the speed at which one can perform the gate while remaining adiabatic; finding  the optimal pulse shape for a given control goal is an area of active research (see, e.g. \cite{martinis2014}). For a sufficiently large energy gap between the dressed ground and excited states, the the adiabatic time scale can be small compare to the time scales for decoherence, such as the finite Rydberg lifetime. In this case, one can remain adiabatic solely by rounding the edges of an essentially square-topped pulse and have minimal impact on gate time.  In the opposite limit, when the energy gap is not very large compared to other decoherence rates, to achieve very high levels of adiabaticity one might require a more triangular pulse, where laser power increases slowly until half the desired phase is accumulated at which point the process is reversed. The parameter ranges we explored fell between these two extremes where adiabaticity was one of a few limiting factors on the gate's speed and fidelity.  An example simulation of the time dependent Schr\"{o}dinger equation in the absence of decoherence is shown in Fig. (\ref{fig:pulseshape}) for the following parameters: pulse rise time 1 $\mu$s, Rabi frequency sweep $\Omega/2\pi =0 \rightarrow$ 3 MHz, detuning sweep $\Delta/2\pi = 6 \rightarrow$ 0 MHz,  Rydberg decay rate $\Gamma/2\pi $= 3.7 kHz, and interatomic separation $\bar{z}=5$ $\mu$m. These parameters produce a blockade shift of $V_{dd}(\bar{z})/2\pi\approx-6.4$ MHz, giving an interaction strength of $J/2\pi\approx 1.8$ MHz at full power. For this example, the populations are highly adiabatic; approximately $99.5\%$ of the original population returns to the ground state.

\section{Motional Errors}
The method described produces a high-fideity CPHASE gate when errors due to motional effects are neglected. To account for the motional degrees of freedom, we must consider the near-degenerate manifold of dressed ground states, all with the same electronic character but different momenta, $\ket{\widetilde{00}}\otimes \ket{p_{rel},P_{cm}}$. The perturbative effects of motion are described by $H_1$, Eq. (\ref{perturbedH}). For a gate performed for atoms in free flight, the finite momentum spread of the atoms leads to two types of errors corresponding to the two terms in $V_{dop}$, Eq. (\ref{motionalVs}). First, the perturbation of the energy, 
\begin{equation}\label{eq:firstorderDop}
\bra{\widetilde{00}}V_{Dop}\ket{\widetilde{00}}=\frac{k_L P_{cm}}{2m}(|c_B|^2+2|c_{rr}|^2),
\end{equation} 
leads to a momentum-dependence of the light shift. This in turn leads to a momentum-dependence of the phase accumulated over the course of the gate, which manifests as decoherence after averaging over motional degrees of freedom. Second, the off-diagonal terms, $\bra{D}V_{dd}\ket{\widetilde{00}}$, transfer population from the ideal dressed ground states into electronic dark states, potentially causing qubit loss as well as decoherence.

An adiabatic gate is naturally robust against some of these motional noise sources. Specifically, the dressed ground manifold is ``protected'' from the excited dressed states by an energy gap, $\Delta E \approx \sqrt{\Delta^2+\Omega^2}$, and by design, we assume that the laser intensity is turned on slowly enough to stay adiabatic given this gap. As long as $ | \bra{e} H_1 \ket{\widetilde{00}} | \ll  |\Delta E|$, averaged over the atomic thermal distribution and for all excited states $\ket{e}$, any time-dependent sweep of the laser parameters that is adiabatic for $H_0$ will also be adiabatic for $H_0+H_1$. Since $H_1$ does not significantly affect adiabaticity, we can completely characterize its effects by examining its action on the dressed ground subspace. By guaranteeing that we remain in a dressed ground state, we make the gate robust against errors that couple the system to states outside the desired 3-level space, $\{\ket{00},\ket{B},\ket{rr}\}$. The off-diagonal bright-dark coupling is such an error, so its effects are largely suppressed. The Doppler shift, on the other hand, is not suppressed and remains a major source of error, even for cold atoms.

\begin{figure*}
[t]\resizebox{12cm}{!}
{\includegraphics{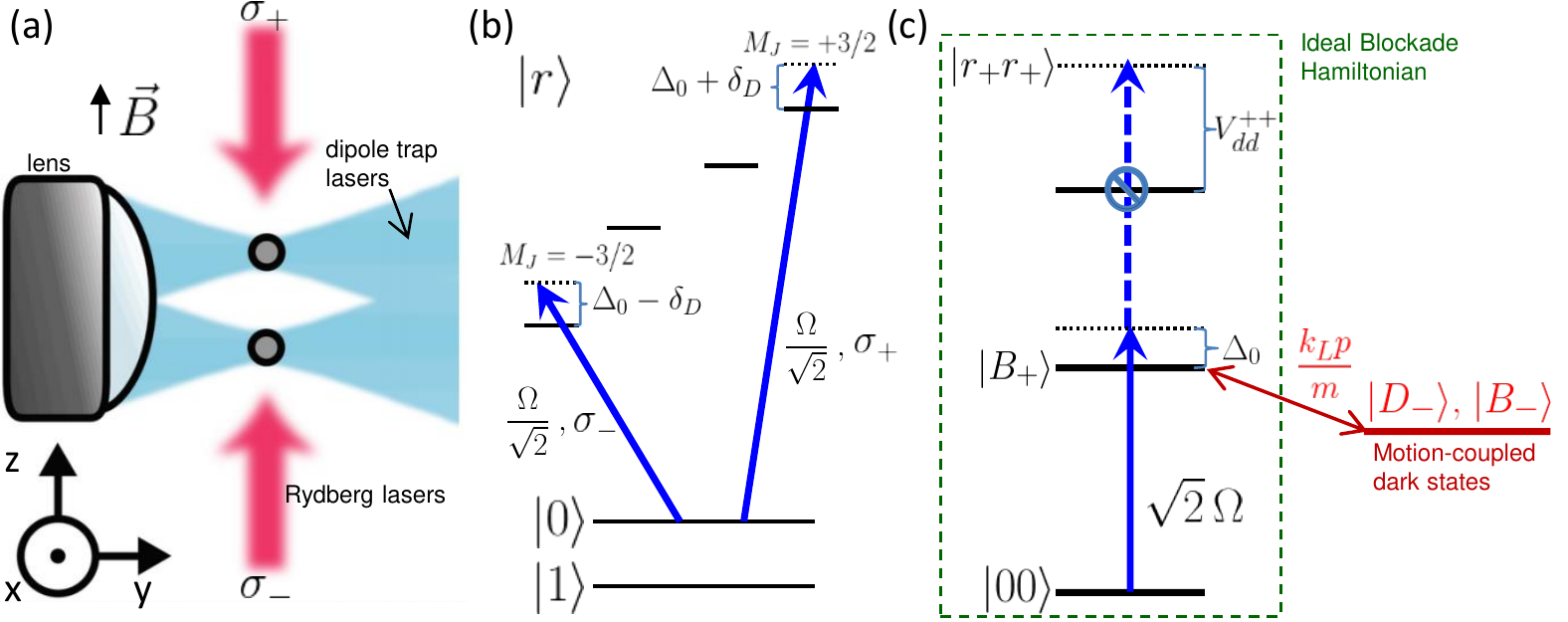}}
\caption{(Color online) (a) Schematic for the ``Doppler-free'' configuration. Two cesium atoms are trapped and cooled in dipole traps, several $\mu$m apart. During the CPHASE gate, the trapping lasers are turned off and the atoms are illuminated by two counterpropagating, 319 nm Rydberg lasers. The two Rydberg lasers have opposite circular polarizations, so they couple the atoms to orthogonal magnetic sublevels of the Rydberg manifold. Both Rydberg lasers propagate along the interatomic separation axis; a bias magnetic field ensures that this coincides with the atomic quantization axis. (b) In each atom, counterpropagating lasers couple the logical-$\ket{0}$ state to the $m_J=\pm\frac{3}{2}$ magnetic sublevels of the $\ket{84P_{3/2}}$ Rydberg manifold. The two lasers have the same Rabi rate $\Omega/\sqrt{2}$ and detuning from resonance $\Delta_0$, but experience opposite Doppler shifts, $\delta_D\equiv k_Lp/m$. Zeeman splitting should be made large enough that coupling to $m_J=\pm\frac{1}{2}$ can be neglected. (c) In the two-atom basis, the states $\ket{00}$, $\ket{B_+}$, and $\ket{r_+r_+}$ are coupled by the ideal blockade Hamiltonian with no Doppler shifts. Instead, motional noise manifests as a coupling to the dark states $\ket{D_-}$ and $\ket{B_-}$. Because $\ket{D_-}$ and $\ket{B_-}$ are outside the ideal adiabatic basis, we can suppress the effects of this coupling through adiabatic evolution.}
\label{fig:level_diagram2}
\end{figure*}

To ensure that Doppler errors are also suppressed, we can make use of a ``Doppler-free'' configuration.  We can achieve this through the addition of the light-shifts from counter-propagating laser-beams on two Rydberg transitions such that the Doppler shift cancels to first order in $p$.  Consider counter-propagating lasers with opposite helicity, $\sigma_+ / \sigma_-$,  tuned to address two different sublevels in the Rydberg manifold (see Fig. \ref{fig:level_diagram2}),
\begin{equation} \label{eq:dopp_free_couplings}
\begin{split}
\sigma_+&: \, \,  \ket{0}= \ket{6S_{1/2}, F=4, m_F=0}\\
& \rightarrow \ket{r_1}= \ket{84P_{3/2}, m_J=3/2}\ket{I=7/2,m_I=-1/2} \\
\sigma_-&: \, \,  \ket{0}= \ket{6S_{1/2}, F=4, m_F=0}\\
& \rightarrow \ket{r_2}= \ket{84P_{3/2}, m_J=-3/2}\ket{I=7/2,m_I=+1/2}
\end{split}
\end{equation}
Note, we choose a  $nP_{3/2}$ Rydberg multiplet because this has  much larger oscillator strength than the corresponding $nP_{1/2}$ mutiplet~\cite{oscillator_strength}. We can suppress the coupling of the $m_F=0$ ground state to the $m_J=\pm1/2$ sublevels with a sufficiently large Zeeman shift so that those transitions remain well off resonance (e.g. $B\approx 10$ G).  Because the two beams are differently detuned and orthogonally polarized, we avoid standing waves in intensity and polarization.

Given the couplings in Eq. (\ref{eq:dopp_free_couplings}), we can write the single-atom Hamiltonian as in Eq. (\ref{eq:1atomH}),
\begin{equation}
\begin{split}
H_A = &\frac{p^2}{2m} - \Delta (\ket{r_1}\bra{r_1}+\ket{r_2}\bra{r_2}) \nonumber \\
&+ \left( \frac{\Omega_1}{2}\, e^{i k_L z}\ket{r_1}\bra{0}+ \frac{\Omega_2}{2} e^{-ik_L z}\ket{r_2}\bra{0} + h.c.\right).
\end{split}
\end{equation}
Including counter-propagating laser beams doubles the incident power, so in order to make a fair comparison to a single laser beam we will assume that $\Omega_1^2 = \Omega_2^2 =  \Omega^2/2$.  In such a configuration, there are coupled and uncoupled excited states for the each of the atoms $ \ket{r_\pm} \equiv  ( \Omega_1\ket{r_1}\pm \Omega_2\ket{r_2})/{\Omega}$.   As before, we can go to a comoving frame, yielding the single atom Hamiltonian
\begin{equation} \label{doppfree1atom}
\begin{split}
H_A &= \frac{p^2}{2m} - \Delta\, \big(\, \ket{r_+}\bra{r_+} + \ket{r_-}\bra{r_-} \, \big)  \\
&+ \frac{k_L\, p}{m}\, \big(\,\ket{r_-}\bra{r_+} + \ket{r_+}\bra{r_-} \, \big) + \frac{\Omega}{2} \big(\,  \ket{r_+}\bra{0} + \ket{0}\bra{r_+}\, \big).
\end{split}
\end{equation}
For this configuration, as in Eq. (\ref{perturbedH}), we can split the two-atom Hamiltonian into $H_0$ for ``frozen atoms'' and a perturbation $H_1$ due to motion.  Thus,
\begin{equation} \label{doppfreeH}
\begin{split}
H_0 &= H_A \otimes \mathbbm{1} + \mathbbm{1}\otimes H_A + V_{dd} \\
  &=  - \Delta(0) \sum_{i = \pm} \left( \ket{B_i}\bra{B_i} + \ket{D_i}\bra{D_i} \right)  \\
 &\quad+ \sum_{i,j = \pm} \left(\, V_{dd}^{i j}(\bar{z}) - 2 \Delta(0) \, \right) \ket{r_i r_j}\bra{r_i r_j} \\
   &\quad + \frac{\sqrt{2}\Omega}{2} \left( \ket{B_+}\bra{00} + \ket{r_+ r_+}\bra{B_+} + h.c.\right)\\
 &\quad + \frac{\Omega}{2} \Big[ (\ket{r_- r_+}+\ket{r_+ r_-})\bra{B_-}\\
&\quad + (\ket{r_- r_+}-\ket{r_+ r_-})\bra{D_-} + h.c.\Big],
\end{split}
\end{equation}
\[
\begin{split}
H_1 =& T+V=\frac{P_{cm}^2}{4m} + \frac{p_{rel}^2}{ m}+ \frac{k_L\, P_{cm}}{2 m} \left( \sigma^{r}_x\otimes \mathbbm{1} +  \mathbbm{1} \otimes \sigma^{r}_x \right)  \\
&\quad+ \frac{k_L\, p_{rel}}{ m} \left( \sigma^{r}_x\otimes \mathbbm{1} -  \mathbbm{1} \otimes \sigma^{r}_x \right)\\
&\quad +\sum_{i,j = \pm}\frac{dV_{dd}^{ij}}{d z}(z-\bar{z})\ket{r_i r_j}\bra{r_i r_j}.
\end{split}
\]
Here, we have defined the Pauli-$x$ operators acting in Rydberg states to be $ \sigma^{(r)}_x \equiv \ket{r_-}\bra{r_+} + \ket{r_+}\bra{r_-}$  as well as the bright and dark states,$\ket{B_\pm} \equiv \left( \ket{r_\pm\,0}\ +\ \ket{0\,r_\pm} \right)/\sqrt{2}$  and  $\ket{D_\pm} \equiv \left( \ket{r_\pm\,0} - \ket{0\,r_\pm} \right)/\sqrt{2}$. The effect of gradient forces now depends in the dipole-dipole potential for the different Rydberg states, $V_{dd}^{ij}(z) = \bra{r_i}V_{dd}(z)\ket{r_j}$.

We see that for the counter-propagating $\sigma_+/\sigma_-$ geometry, $H_0$ is block diagonal in the electronic degrees of freedom as well as diagonal in $p$. The states $\ket{00}$, $\ket{B_+}$, and $\ket{r_+ r_+}$ form a block described by our desired 3-level blockade Hamiltonian, while $\ket{B_-}$, $\ket{D_-}$, $\ket{r_+,r_-}$, and $\ket{r_-,r_+}$ form a separate block; the state $\ket{D_+}$ is completely uncoupled from all other states. The terms in $V$ arising from the Doppler shift scale as $k_L p \sigma_x^r/m$, but because this coupling is off-diagonal, its effect will manifest as a second order perturbation to the energies of $\ket{B_+}$ and $\ket{r_+ r_+}$. This counter-propagating laser configuration can thus be considered as ``Doppler-free'' to first order.  By contrast, with a single laser beam, $\bra{B}V\ket{B}$ was nonzero, leading to contributions to the dressing energy that are first order in the Doppler shift.  To zeroth order in $p$, our scheme only involves the states in the $3\times3$ ideal block; the other states are only included through perturbations.  Restricting $H_0$ to this subspace leaves
\begin{equation}
\begin{split}
  H_0 &=  V_{dd}^{++}(z) \ket{r_+ r_+}\bra{r_+ r_+}  - \Delta \Big(\ket{B_+}\bra{B_+} + 2 \ket{r_+ r_+}\bra{r_+ r_+} \Big)\\
   &+\frac{\sqrt{2}|\Omega|}{2} \Big( \ket{B_+}\bra{00} + \ket{00}\bra{B_+} +  \ket{r_+ r_+}\bra{B_+} + \ket{B_+}\bra{r_+ r_+} \Big),\\
\end{split}
\end{equation}
a Doppler-free Hamiltonian (see Fig. \ref{fig:level_diagram}).

The ability to suppress motional error via this Doppler-free configuration is a key benefit of the adiabatic gate approach. For comparison, consider the effects of the same error Hamiltonians on a gate protocol based on fast pulses \cite{saffmanreview}. Such a gate involves the application of resonant lasers on one atom at a time in a series of unitary evolutions: a $\pi$-pulse excites a control qubit in one logical state to the $\ket{r}$ state followed by a $2 \pi$-pulse  applied to the target qubit; the control qubit is then de-excited by another $\pi$-pulse.  During its time $T=2 \pi/\Omega$ in the Rydberg state, the control qubit freely evolves, resulting in a phase accumulation due to the Doppler shift, $\exp(-2 \pi i\frac{k_Lp}{m\Omega})$.  This error is first-order in $p$, as in the single-laser adiabatic protocol.  Using the counter-propagating $\sigma_+/\sigma_-$ laser geometry, the situation is similar, except that now each atom evolves according to the Hamiltonian $H_A$, Eq. (\ref{doppfree1atom}). During the time $T$ the off-diagonal terms of the Hamiltonian cause the control qubit to evolve from $\ket{r_+}$ to $\cos(2 \pi \frac{k_Lp}{m\Omega})\ket{r_+}+\sin(2 \pi\frac{k_Lp}{m\Omega})\ket{r_-}$. Any population transferred to $\ket{r_-}$ will be uncoupled from the de-exciting $\pi$-pulse, and this leads loss of probability amplitude that is first-order in $p$.  The fast pulse scheme cannot be made ``Doppler-free'' to first order. In contrast, adiabatic evolution suppresses population transfer to states outside the $3\times3$ ideal block, so this population loss is greatly reduced; it only manifests as a second-order energy perturbation, which leads to errors a factor of $\sim\frac{k_Lp}{m\Omega}$ smaller.

In addition to the effect of finite momentum spread, recent work has shown that the Rydberg interaction itself can lead to further two-body decoherence when the blockade is imperfect \cite{nottingham}.  Because the dipole-dipole energy $V_{dd}$ varies with interatomic distance, it can produce an interatomic force when the system is in $\ket{rr}$.  In our case, the effect of the force is captured by ${V_{grad}}$, Eq. (\ref{motionalVs}), which does not change in the Doppler-free geometry.  The perturbation on the dressed ground state is $\bra{\widetilde{00}}V_{grad}\ket{\widetilde{00}} =|c_{rr}|^2 \frac{dV_{dd}}{dz}(z-\bar{z})$, leads to a displacement on the relative momentum of atoms in this state, $\delta p_{rel}=\displaystyle\int_0^T|c_{rr}(t)|^2\frac{dV_{dd}}{dz} dt$.  Higher order perturbations take the system out of its dressed ground state to some excited state $\ket{e}$; as long as the evolution remains adiabatic, they are suppressed by an extra order of $|\bra{e}V_{grad}\ket{\widetilde{00}}|/\Delta E$.  For a near ``perfect blockade,''  where $|V_{dd}|\gg\Delta, \Omega$, and $c_{rr}\approx0$, this force can be neglected entirely.

\section{Simulated Gate Fidelities}

To evaluate the performance of the gate, we use as our metric the fidelity to produce the desired output given an input of all the logical states, $\ket{\psi_0}= \left(u_{H} \otimes u_{H} \right)\ket{00}$, where $u_{H}$ is the Hadamard gate. This fidelity $\mathcal{F}= \bra{\psi_{tar}}\rho_{out} \ket{\psi_{tar}}$, where $\ket{\psi_{tar}}$ is the target state obtained through a combination of local unitaries and an ideal CZ gate, $\ket{\psi_{tar}}= U_{CZ}\ket{\psi_0}  =\frac{1}{2} \left( \ket{11}+\ket{10}+\ket{01}-\ket{00}\right)$, while $\rho_{out}$ is the actual state in the logical space produced in the presence of the error sources described above: nonadiabatic dressing, decay of the Rydberg state, Doppler shift, and dipole-dipole forces for an imperfect blockade,
\begin{equation} \label{eq:rho_out}
\begin{split}
\rho_{out} =& {\rm Tr}_{\rm ext} \Big[ e^{-i\ket{00}\bra{00}\otimes\delta p_{rel}z} U_{\rm eff}\left( \ket{\psi_0}\bra{\psi_0}\otimes \rho^{\rm ext} \right) U^\dag_{\rm eff}\\
 &\quad\times e^{i\ket{00}\bra{00}\otimes\delta p_{rel}z} \Big ].
\end{split}
\end{equation}
Here $\rho^{\rm{ext}}$ is the thermal state associated with the ``external'' (motional) degrees of freedom, $\delta p_{rel}$ is the total momentum displacement caused by the dipole force, and $U_{\rm{eff}}$ is the total effective action of the gate including all decoherence sources other than the dipole-dipole force.  It is nonunitary due to the non-Hermitian Hamiltonian arising from decay of the Rydberg state and thus we treat the map as generally non-trace-preserving. We are able to separate out the effects of the dipole force through a first-order Baker-Campbell-Hausdorff expansion; since $H_0$ commutes with momentum displacements, all higher-order terms will scale as the products of already small error Hamiltonians and can be ignored.  Because $U_{\rm eff}$ does not couple different logical states, it is convenient to expand $\mathcal{F}$ in the logical basis, giving
\begin{equation} \label{eq:three_fidelity_terms}
\mathcal{F}=\frac{1}{4}\displaystyle\sum_{x,y,x',y'}\left(-1\right)^{\delta_{xy,00}-\delta_{x'y',00}}\bra{xy}\rho_{out}\ket{x'y'}
\end{equation}
where $\ket{xy}$ are over the two-qubit logical states.

\begin{figure}
\includegraphics{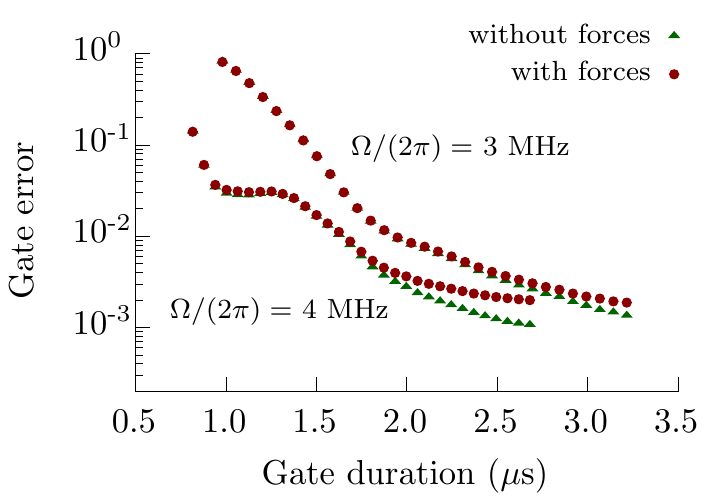}
\caption{\label{fig:BlockadeComparison} (Color online) Simulated gate error rates ($1-\mathcal{F}$) as a function of adiabatic ramp time. The upper pair of curves were generated with the parameters given in Fig.~{\ref{fig:pulseshape}}, while the lower curves used a higher Rabi rate for the exciting laser. Ignoring interatomic forces but including all other errors (green triangles), the higher Rabi rate improves both gate speed and fidelity. Including interatomic forces (red circles), any gain in fidelity from the increased speed is offset by stronger forces owing to a larger $\ket{rr}$ population when the blockade is imperfect. This suggests that beyond a certain threshold, increased laser power requires a commensurately stronger blockade interaction in order to improve fidelity.}
\end{figure}

To understand the effects of atomic motion on gate errors, consider the  contribution to the fidelity from each of the matrix elements in Eq. (\ref{eq:three_fidelity_terms}) under the assumption of perfect adiabatic evolution of the dressed states.  When both atoms are in the logical-$1$ state, we assume no coupling to the laser, and thus there is no error contribution from $\bra{11}\rho_{out}\ket{11}$.  When both atoms are in the logical-$0$ state, both photon scattering and motional effects come into play. Motional dephasing has no effect on populations, only photon scattering contributes error on the diagonal terms of $\rho_{out}$,
\begin{equation}
\bra{\widetilde{00}}\rho_{out}\ket{\widetilde{00}}=\frac{1}{4}e^{-\gamma T},
\end{equation}
where the factor $e^{-\gamma T}$ accounts for loss due to the finite lifetime of the Rydberg state $\gamma T =\Gamma \int_0^T \left(|c_B (t')|^2+2 |c_{rr} (t')|^2\right) dt'$. On the other hand, the off-diagonal terms are affected by both loss and dephasing, leaving
\begin{equation} \label{eq:doppler_matrix_elements}
\begin{split}
\bra{11}&\rho_{out}\ket{\widetilde{00}} = -\frac{1}{4} e^{-\gamma T/2} \int dP_{cm} dp_{rel} \, e^{-i \phi_{Dop}}\\
&\quad\quad\times \bra{P_{cm},p_{rel} }\rho^{ext} \ket{P_{cm},p_{rel}}\\
& =  -\frac{1}{4} \, e^{-\gamma T/2} \int dP_{cm} dp_{rel}  \, e^{-i \phi_{Dop}}   \frac{e^{-\frac{P^2_{cm}}{4 \Delta p_{th}^2}} e^{-\frac{p^2_{rel}}{\Delta p_{th}^2}}}{4 \pi \Delta p_{th}^2}.
\end{split}
\end{equation}
We have assumed a thermal state of motion associated with the initial trapped atom of mass $m$ with mean vibrational quantum number $\bar{n}$, with $ \Delta p_{th}^2 = (\bar{n}+1/2)m\omega_{osc}$, and used the fact that the Doppler effect is diagonal in the momentum representation.  The additional phase, $e^{-i \phi_{Dop}}$, is due to perturbation of the dressed ground state energy arising from the Doppler shift,
\begin{equation}
\begin{split}
\phi_{Dop} (P_{cm},p_{rel}) \equiv\displaystyle\int_0^T \Bigg(\bra{\widetilde{00}(t')}V_{Dop}\ket{\widetilde{00}(t')}\\
+\displaystyle\sum_e \frac{\left|\bra{e}V_{Dop} \ket{\widetilde{00}(t')}\right|^2}{\bra{\widetilde{00}(t')}H_0\ket{\widetilde{00}(t')}-\bra{e}H_0\ket{e}} \Bigg) dt'.
\end{split}
\end{equation}
With a single coupling laser the correction to the light shift, Eq. (\ref{eq:firstorderDop}), is first order in $p$, and Eq. (\ref{eq:doppler_matrix_elements}) can be integrated analytically. This leads to a reduction in the fidelity of order $e^{-(\bar{n}+1/2) \eta^2 (\omega_{osc} T)^2}$, where $\eta =\sqrt{E_{recoil}/\hbar \omega_{osc}}$ is the Lamb-Dicke parameter.  For example, using the parameters in Fig. \ref{fig:pulseshape} and $\bar{n}=5$, we find that the $\bra{11}\rho_{out}\ket{00}$ coherence is reduced to $\sim0.90$ of its original value due to Doppler effects - an order of magnitude more decoherence than from any other source. In contrast, with the Doppler-free configuration, the first order correction vanishes, thereby strongly suppressing the effect of the Doppler shift. The $\ket{01}$ and $\ket{10}$ states experience similar Doppler perturbations to their single atom light shifts, which are generally different from the light shifts on $\ket{00}$. This means that the coherences between $\{\ket{01},\ket{10}\}$ and $\{\ket{11},\ket{00}\}$ are also significantly reduced by Doppler effects, and the Doppler-free configuration likewise suppresses these decoherences.

The effect of the dipole-dipole force is seen in the coherences $\bra{xy}\rho_{out}\ket{00}$, where $xy \neq 00$.  Because atoms in $\ket{00}$ will experience a relative momentum kick when the blockade is imperfect and they are both excited into the Rydberg state, this local basis state will contain ``which way'' information relative to the other basis states.  Tracing over the the motional degrees of freedom, this leads to a reduction of the coherences, 
\begin{equation} 
\label{eq:dipole_force_decoherence}
\begin{split}
&\bra{xy}\rho_{out}\ket{00} \propto {\rm Tr}_{\rm ext} \left[ e^{-i\delta p_{rel}z}  \rho^{\rm ext}_{rel} \right]\\
&\quad= \int dp_{rel} \, \bra{p_{rel} + \delta p_{rel}} \rho^{\rm ext}_{rel} \ket{p_{rel}} =e^{-\frac{(\bar{n}+1/2)\delta p_{rel}^2}{2 M \omega}}.
\end{split}
\end{equation}
Because $\delta p_{rel}$ scales with $\ket{rr}$ population, this decoherence provides a strong penalty for increasing the exciting laser power beyond the point of ``breaking'' the blockade (see Fig. \ref{fig:BlockadeComparison}). For this reason, strong blockade interactions as well as high Rabi rates will be required to achieve very high fidelities.

Finally, the gate's fidelity is reduced by imperfect adiabatic following. Diabatic transitions during dressing process generally cause both population loss and dephasing for each atom in the $\ket{0}$ state, so nearly every element of $\rho_{out}$ is affected. The magnitude of the resulting fidelity loss can be found by numerical simulation.

To calculate the fidelity according to Eq. (\ref{eq:three_fidelity_terms}), we simulate the evolution according to the (non-Hermitian) time-dependent Schr\"{o}dinger equation governed by $H_{eff}$.  This generates the (non trace preserving) evolution $U_{eff}$, accounting for errors due to imperfect adiabatic evolution, loss of atoms due to excitation to the Rydberg state, and decoherence due to thermal spread of Doppler shifts.  We use the simulated excitation to $\ket{rr}$ to calculate the relative momentum kick given to atoms due to the dipole-dipole force, and from this include the additional decoherence effect described in Eq. (\ref{eq:dipole_force_decoherence}).

\begin{figure}
\includegraphics{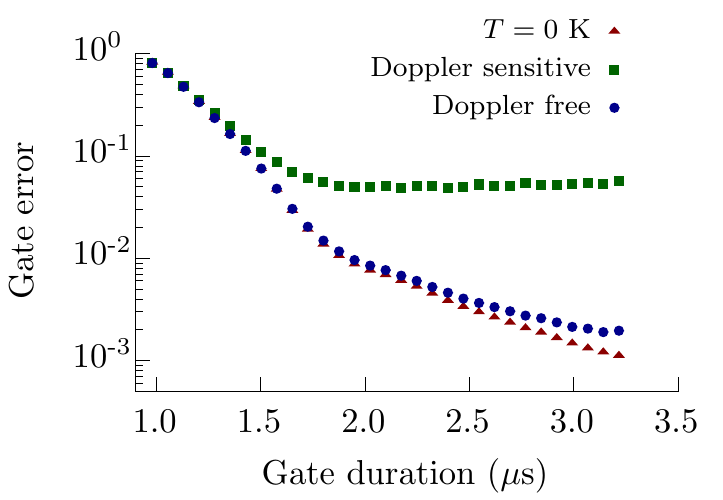}
\caption{\label{fig:FidelityComparison} (Color online) Simulated gate error rates ($1-\mathcal{F}$) as a function of adiabatic ramp time. For comparison, the red triangle curve ignores motional effects and includes errors due solely to diabatic transitions and finite Rydberg lifetime. For ramp times significantly below $1$ $\mu$s, all curves predict  low  fidelities because the gate is not adiabatic. As the ramp time and adiabaticity are increased, other error sources become limiting factors. Including all error sources while using the Doppler-free configuration (blue circles), we can reach error rates of $\sim2 \times 10^{-3}$, with finite blockade strength as the primary fidelity-limiting factor. By contrast, the single-laser configuration (green squares) suffers more than an order of magnitude greater error than its counterparts.}
\end{figure}

As an example, we take the parameters given in Fig.~{\ref{fig:pulseshape}}.  This requires a ramp time on the order of $1$ $\mu$s to stay adiabatic, so that one can perform a CPHASE gate in $\sim2.3$ $\mu$s.  Putting together all of the error sources discussed, we calculate a gate infidelity of $1-\mathcal{F}\sim 2\times 10^{-3}$ for the Doppler-free configuration.  The gate error arises in small part from the second-order effect of Doppler shifts and finite Rydberg lifetime, but it is dominated by interatomic dipole forces owing to an imperfect blockade (see Fig. \ref{fig:FidelityComparison}).  By contrast, without the Doppler-free configuration, the same parameters give an infidelity of $1-\mathcal{F}\sim.04$, almost all of which is due to the spread in Doppler shifts.

\section{Conclusion}
We have studied a method for robustly implementing a CZ gate between neutral cesium atoms based on adiabatic dressing of the ground state via the Rydberg blockade.   The main advantage of this approach is that it strongly suppresses random phases between bright and dark-state superpositions that arise due to atomic motion.  In addition, by employing two counterpropagating Rydberg lasers in a $\sigma_+/\sigma_-$ configuration, one can eliminate the Doppler shift to first order. All effects of thermal motion then take the form of coupling to a dark state outside the ideal blockade subspace, which is suppressed by an energy gap during adiabatic evolution.  When both adiabatic dressing and the Doppler-free configuration are used together, errors from thermal motion are reduced by more than an order of magnitude compared to either strategy used alone.

With motional errors reduced in this way, the main remaining source of error is entanglement between internal and external degrees of freedom due to dipole-dipole forces when the Rydberg blockade is imperfect. Such error is highly nonlinear in laser power; it can be kept small as long as the Rydberg blockade is nearly perfect, but increases rapidly when laser power is increased beyond the point of breaking the blockade. This implies that the available blockade strength sets an upper limit on useful laser power, which in turn limits both the fidelity and speed of the gate.  If the blockade shift can be increased by bringing atoms into closer proximity or by the appropriate choice of Rydberg levels, the gate errors will be limited solely by finite Rydberg lifetime.

As a final note, we have considered here gates performed while atoms are untrapped and fall ballistically.  Recapturing the atoms after the gate will generally cause the atoms to heat~\cite{pfau}.  This effect is not reflected in our error estimates because it does not affect the fidelity of any one gate, but it could increase decoherence if multiple gates are performed successively with no re-cooling in between.  In principle, all of these errors would be substantially reduced in a ``magic trap'' which traps electronic-ground-state and Rydberg atoms equivalently~\cite{Topcu2013}.  In that case, cooling the atoms to the vibrational ground state would completely remove Doppler shifts as well as suppress decoherence due to the dipole-dipole force in an imperfect blockade, providing a potential path to high-fidelity quantum logic.

\vspace{0.5 cm}

\emph{Acknowledgments:} This work was supported by the Laboratory Directed Research and Development program at Sandia National Laboratories.  Sandia National Laboratories is a multi-program laboratory managed and operated by Sandia Corporation, a wholly owned subsidiary of Lockheed Martin Corporation, for the U.S. Department of Energy's  National Nuclear Security Administration under contract  DE-AC04-94AL85000.

\bibliography{CPHASEbib}

\end{document}